\documentclass[11pt,nofootinbib]{revtex4-1}

\usepackage{amsmath,amssymb}
\pdfoutput=1

\usepackage{graphics}
\usepackage[dvips]{graphicx}
\usepackage{mathrsfs}
\usepackage{amssymb}
\usepackage{amsmath}
\usepackage{verbatim}
\usepackage{float}
\usepackage{slashed}
\usepackage{bbm}
\usepackage[dvips,letterpaper,text={6.5in,9in}]{geometry}
\newcommand{\mttwo}{m_{T2}}
\newcommand{\mll}{m_{\ell\ell}}
\newcommand{\slep}{\tilde{l}}
\newcommand{\be}{\begin{equation}}
\newcommand{\ee}{\end{equation}}
\newcommand{\bea}{\begin{eqnarray}}
\newcommand{\eea}{\end{eqnarray}}

\newcommand{\gev}{\text{~GeV}}

\newcommand{\met}{\slashed E_T}
\newcommand{\none}{\chi^0_1}
\newcommand{\ntwo}{\chi^0_2}

\begin{document}
\setcounter{page}{0}

\title{$M_{T2}$ to the Rescue -- Searching for Sleptons in Compressed Spectra at the LHC}
\author{Zhenyu Han}
\affiliation{\small \sl Institute for Theoretical Science, University of Oregon, Eugene, OR 97403, USA}
\author{Yandong Liu}
\affiliation{Department of Physics and State Key Laboratory of Nuclear Physics and Technology,
Peking University, Beijing 100871, China}

\begin{abstract}
We propose a novel method for probing sleptons in compressed spectra at hadron colliders. The process under study is slepton pair production in $R$-parity conserving supersymmetry, where the slepton decays to a neutralino LSP of mass close to the slepton mass. In order to pass the trigger and obtain large missing energy, an energetic mono-jet is required. Both leptons need to be detected in order to suppress large standard model backgrounds with one charged lepton. We study variables that can be used to distinguish the signal from the remaining major backgrounds, which include $t\bar t$, $WW$+jet, $Z$+jet, and single top production. We find that the dilepton $m_{T2}$, bound by the mass difference, can be used as an upper bound to efficiently reduce the backgrounds. It is estimated that sleptons with masses up to about $150 \gev$ can be discovered at the 14 TeV LHC with 100 fb${}^{-1}$ integrated luminosity.
\end{abstract}

\maketitle

\section{Introduction}
\label{sec:intro}
Low energy supersymmetry (SUSY) is an attractive theory of physics beyond the standard model (SM). In order to avoid fine tuning to the Higgs mass, super partners of the SM particles are predicted to be around or below the TeV scale, which is often dubbed ``natural supersymmetry'' -- see Ref.~\cite{Kribs:2013lua} and references therein. However, SUSY searches at the large hadron collider (LHC) have not revealed any signal beyond the standard model, which have put stringent constraints on the SUSY mass spectrum. To reconcile the null results with supersymmetry, one either (partially) gives up naturalness and accepts that the super particles' masses are beyond the current reach of the 8 TeV LHC (which could, however, be discovered at 14 TeV or a future collider), or assumes SUSY particles are light and accessible, but the signal is hidden in the SM backgrounds. In order not to miss the SUSY signals, both the two possibilities should be explored. One way to hide light SUSY particles is to make the spectrum compressed, that is, the mass splittings among the SUSY particles are so small that the decay products of the SUSY cascades are soft. The signal events that contain such soft particles, including jets, leptons or photons, are difficult to trigger on, and even if recorded, they are usually buried in SM backgrounds. Special search strategies are required to find the signal events and previous studies include those on a light stop \cite{Han:2012fw, Alves:2012ft, Kilic:2012kw, Bai:2013ema}, a light sbottom \cite{Alvarez:2012wf} and light electroweakinos \cite{Gori:2013ala, Schwaller:2013baa, Baer:2014cua, higgsino,  Han:2014xoa, Baer:2014kya}. In this article, we focus on another important SUSY process, slepton pair production. 

We assume the lightest supersymmetric particle (LSP) is a neutralino with mass around 100 GeV. A light slepton with mass close to the LSP mass is not required by naturalness because its loop contribution to the Higgs mass is small. Nevertheless a $5\sim 20\gev$ mass splitting, which we assume in this article, is certainly possible without ``fine-tuning'' model parameters. Moreover, such a small splitting is needed to obtain the correct relic density in the co-annihilation scenario \cite{Griest:1990kh}. When sleptons are pair produced and each of which decays to a neutralino, we have two soft leptons and missing energy in a signal event. The major SM backgrounds include $t\bar t$, $WW$+jet, $Z(\rightarrow)\tau\tau$+jet and single top production. In order to pass the trigger, we require an extra hard jet and large missing energy to be present in the event. This is also the final state particles considered in Refs.~\cite{higgsino, Baer:2014kya}, where the discovery potential of the LHC for quasi-degenerate Higgsinos is explored.
A crucial observation in the analysis which makes the discovery possible is the fact that the majority of the lepton pairs are produced through off-shell $Z's$ in $\ntwo \rightarrow \none$ decays, and the dilepton invariant mass $\mll$ is bound from above by the  $\ntwo-\none$ mass difference. Therefore, we can apply an upper cut on $\mll$ to eliminate bulk of the background events, while retaining most of the signal events. This feature is unfortunately absent for slepton pair production because the two leptons necessarily come from two different decay chains. For a typical 10 GeV lepton $p_T$ acceptance cut, the dilepton invariant mass spreads from $\sim 10\gev$ to $\sim 80\gev$, which significantly overlaps with the SM backgrounds. Clearly, a different strategy is needed.

In this article, we propose a novel method for searching slepton pairs in a compressed spectrum. In order to exploit the small mass splitting, we consider the $\mttwo$ variable defined from the two leptons and the missing transverse momentum. This variable, to a good approximation, is bound by the mass difference between the slepton and the LSP. Because of this property, we use it as an {\it upper} bound in our method. This is in contrast to the traditional use of $m_{T2}$ in SUSY searches, where $\mttwo$ is a variable alternative to the missing transverse momentum and usually used as a lower cut to reduce SM backgrounds. As we will show, this variable is the most efficient among known variables that are sensitive to the small mass splitting and can distinguish the signal from the SM backgrounds. We perform an analysis for the 14 TeV LHC: assuming an integrated luminosity of 100 fb${}^{-1}$, the signal can be discovered up to $150\gev$ for left-handed sleptons.  

The rest of the paper is organized as follows. We describe our method and simulation details in Section \ref{sec:analysis}. The LHC discovery limits are presented in Section \ref{sec:results}. Section \ref{sec:conclusion} contains some discussions and we compare to a few other variables in the Appendix.

\section{Simulations and analysis}
\label{sec:analysis}
The signal considered in this article has a simple event topology. A pair of sleptons is produced from $Z$ or $\gamma$ exchange from a pair of initial state quarks. Each slepton then decays to a neutralino LSP and a lepton. When the slepton mass is larger than the neutralino mass by only a small amount, $\sim10\gev$, we face two difficulties when trying to detect the signal. First, the signal event contains only soft particles and a small missing energy. Therefore, it usually does not pass the trigger and the event is lost. Second, even if the event is recorded, the acceptance for the soft leptons is low, one or both of the two leptons are often lost. Because of these difficulties, searches for dilepton + missing energy did not reach this region of the parameter space. In the latest LHC results, no constraint is set when the mass difference is below $\sim60\gev$ \cite{Khachatryan:2014qwa, Aad:2014vma}.

We can alleviate the two difficulties by requiring an extra hard jet to be present: it provides a monojet plus missing energy signature for the event to be trigged on, and gives the slepton pair a boost to increase their $p_T$'s. Due to the low acceptance of soft leptons, we then need to decide how many leptons in addition to the monojet have to be detected for the search. As discussed in Ref.~\cite{higgsino}, the monojet signal alone will not provide a more stringent bound than LEP 2 for degenerate Higgsinos. Being an electroweak process, the slepton pair cross section is similar to that of Higgsino pairs. Therefore, the conclusion still holds and we do not expect a better bound from the LHC than LEP 2 \cite{slepton-lep}, which is around $100\gev$. It is also challenging to consider events with only one lepton detected. The background from the SM $W$+jet contains the same visible particles and the cross section is enormous. The fake rate for one lepton is also much higher than for two leptons. Therefore, in this article, we will require both leptons to be accepted by the detector, and the event is characterized by one energetic jet, significant amount of missing energy and two leptons. Even with this requirement, the SM backgrounds are still overwhelming which requires special techniques.

As discussed in Ref.~\cite{higgsino}, the major SM backgrounds that contain two isolated leptons include $t\bar t$, $\ell\ell\nu\nu$ +jet (dominated by $WW$+jet) and $Z(\rightarrow \tau\tau)$+jet, in the dileptonic channels. Fake leptons, either from light flavor jets faking leptons in $W$+jets, or from heavy flavor decays in $Wb\bar b$, are much smaller than the major backgrounds. On the other hand, as pointed out in Ref.~\cite{Baer:2014kya}, single top production is another background that needs to be included in the analysis. Single top production has a large cross section, only a factor of $\sim 3$ smaller than $t\bar t$. In a single top event, we get an isolated lepton and significant missing $E_T$ when the $W$ decays leptonically. The other lepton is obtained from one of the $b$-hadron decays. As we will see, this background is sizable, but smaller than other backgrounds after all cuts are applied. One may also be concerned about the background from $t\bar t$ semileptonic decays, which can also yield 2 leptons and missing energy. However, as we will veto a second hard jet, the presence of 4 hard QCD partons in a semileptonic $t\bar t$ event makes it very difficult to pass the cut, and the background turns out to be negligibly small. In summary, we will include in our analysis $t\bar t$, $\ell\ell\nu\nu$+jet, $Z$+jet in their dileptonic decay channels, and single tops.

Signal and background processes are generated for the 14 TeV LHC with Madgraph 5 \cite{Alwall:2014hca}, which are then processed with Pythia 6 \cite{Sjostrand:2006za} for showering and hadronization. We quote results using the leading order cross sections given in Madgraph. The leading order sections are typically smaller than the NLO results. Given that electroweak processes associated with an extra hard jet may have a large $k$-factor, $\sim 2$, \footnote{For neutralino pair production associated with a jet, Ref.~\cite{Cullen:2012eh} gives a $k$-factor of 2.3. We expect a similar $k$-factor for slepton pair production because the process involves the same initial states.} the tree level result is a conservative estimate. In order to take into account experimental resolutions, we use Delphes 3 \cite{deFavereau:2013fsa} for fast detector simulations. We use the default Delphes 3 run card in our simulation except for two modifications. First, the default acceptance threshold for leptons is 10 GeV. For very small mass splittings ($\sim 5\gev$), decreasing the threshold will significantly increase the signal efficiency. Therefore, we have set it to be 7 GeV, which is comparable to the threshold used by ATLAS/CMS \cite{ATLAS:2012ac, ATLAS:2013nma, Chatrchyan:2012ufa}. The efficiencies for identifying leptons are set to be 0.95 for muons with $|\eta|\leq2.4$, and 0.95 (0.85) for electrons with $|\eta| \leq 1.5$ ($1.5<|\eta| < 2.5$). Second, we have modified the b-tagging efficiency to 0.7 for jets satisfying $p_T>20\gev$ and $|\eta| < 2.5$ (and 0 for jets not within these limits), as one of the bench mark values used by ATLAS/CMS \cite{ATLAS:2011qia, Chatrchyan:2012jua}. Since the largest background is from $t\bar t$, a high b-tagging efficiency is crucial for vetoing events containing b-jets and reducing this background. For this reason, a more aggressive b-tagging efficiency is preferred. For example, in Ref.~\cite{Chatrchyan:2012jua}, it is shown that a b-tagging efficiency of 0.85 is achieved when the fake rate for light jets is 0.1. Comparing with the value 0.7 we use, we would reduce the $t\bar t$ background by a factor of $\sim 2$, while only losing 10\% of the signal events.
 
\begin{figure}
\begin{center}
\includegraphics[width=0.7\textwidth]{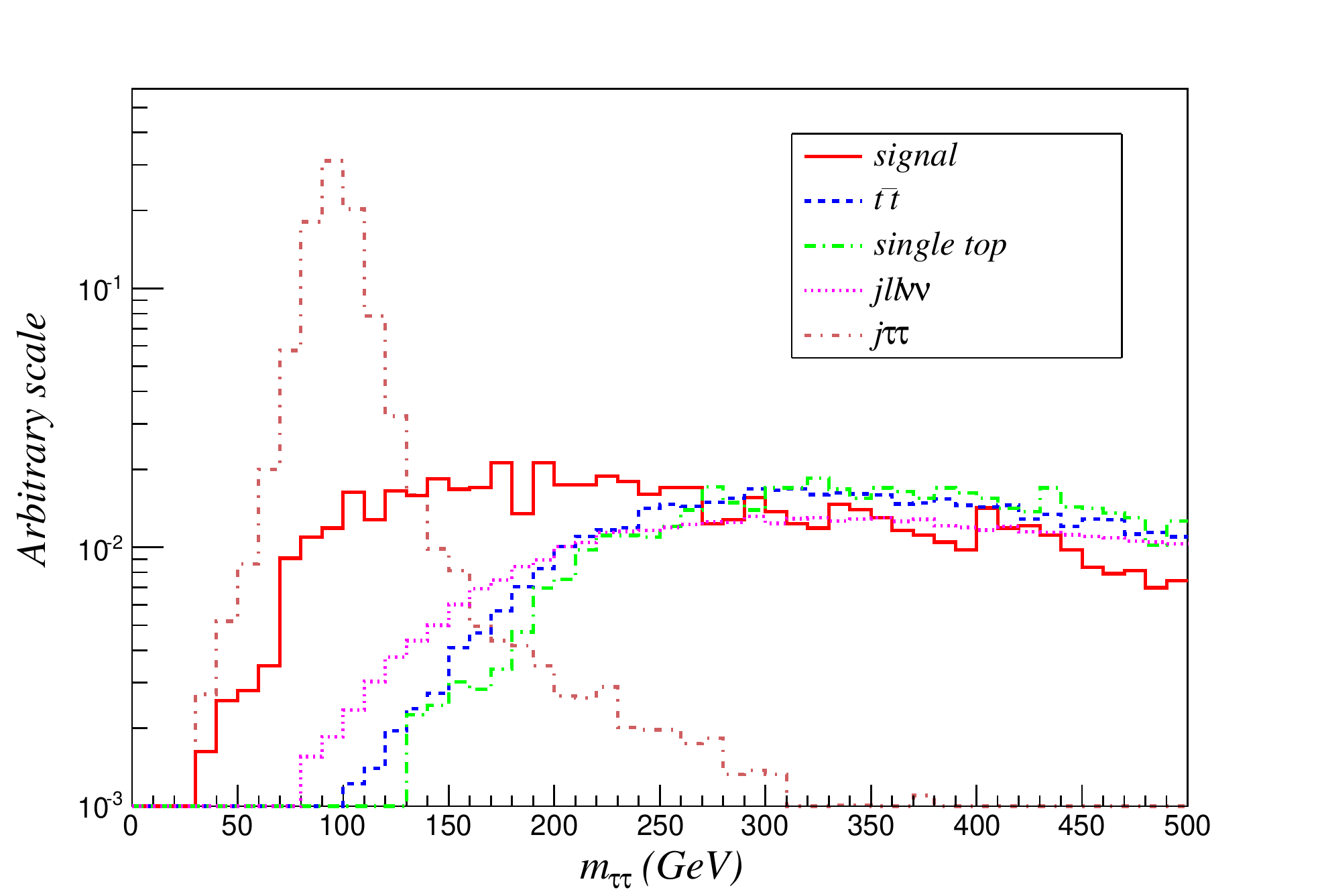} 
\end{center}
\caption{The reconstructed $m_{\tau\tau}$ distributions, normalized to the same area. Events included in this figure have passed cuts 1-4. }
\label{fig:mtautau}
\end{figure}

\begin{figure}
\begin{center}
\begin{tabular}{cc}
  \includegraphics[width=0.5\textwidth]{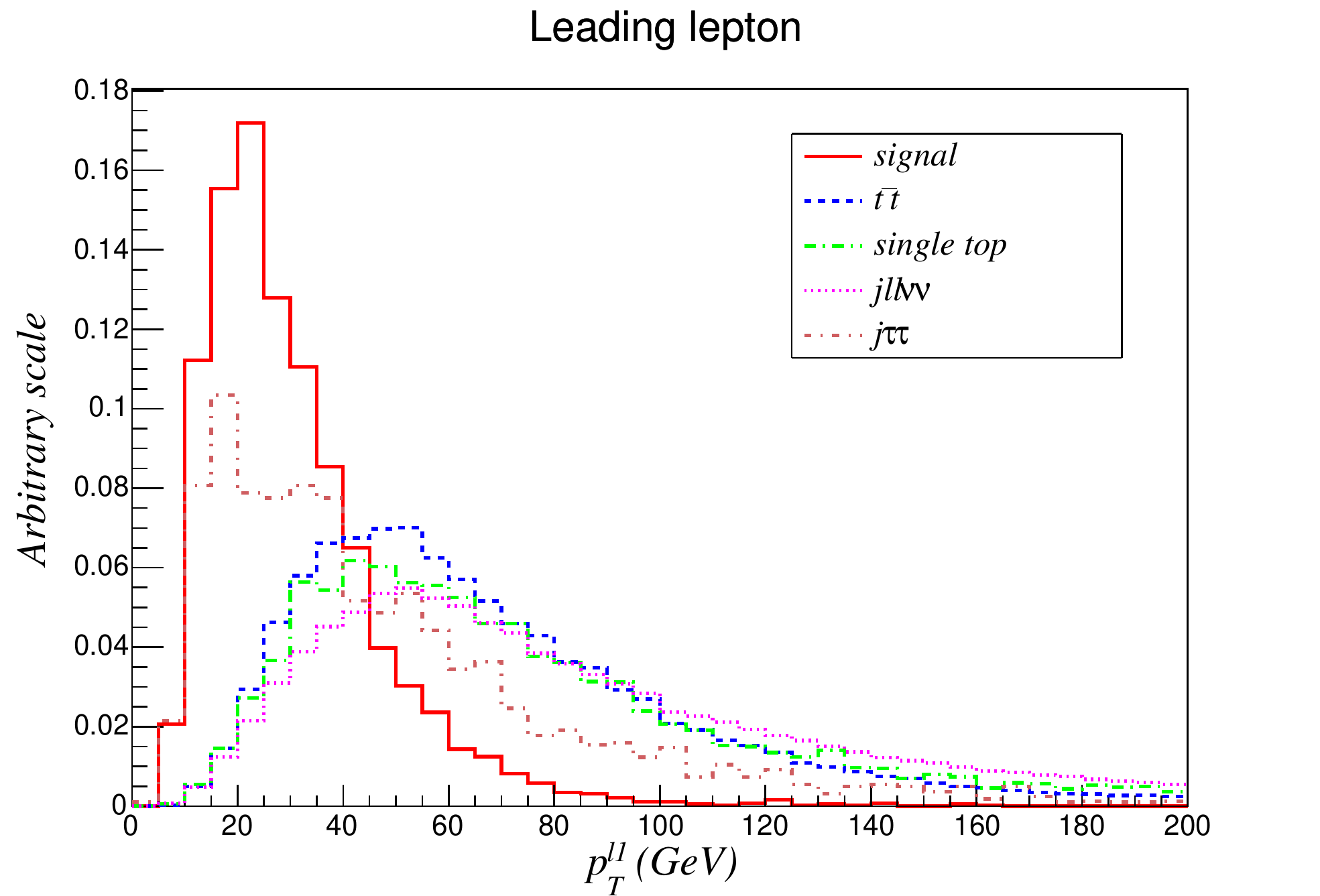} &
  \includegraphics[width=0.5\textwidth]{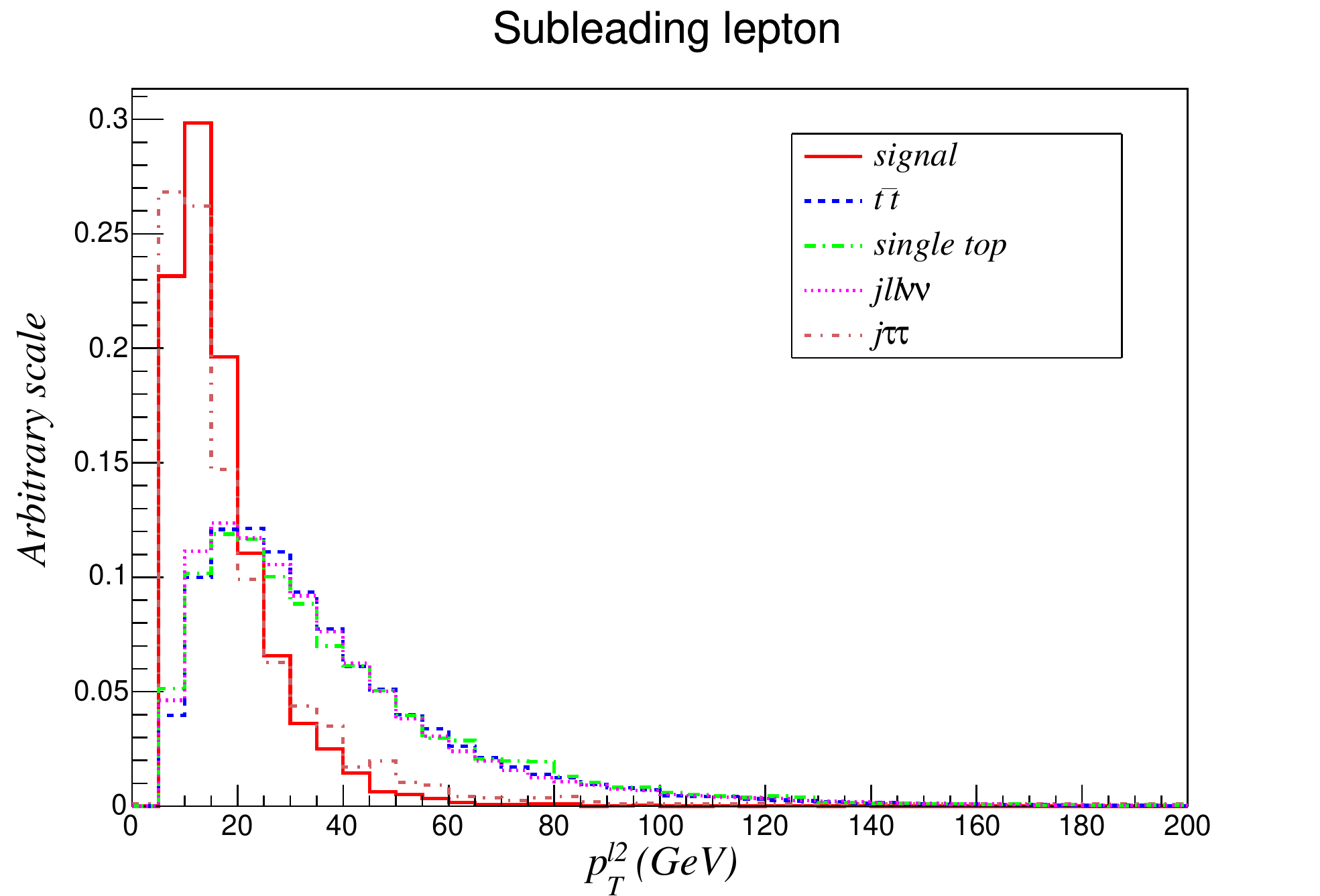}
\end{tabular}  
\end{center}    
\caption{The $p_T$ distributions for the leading and subleading leptons, normalized to the same area.  Events included in this figure have passed cuts 1-5.}
\label{fig:pt}
\end{figure}

We use the following kinematic cuts to reduce the backgrounds, some of which are similar to Ref.~\cite{higgsino, Alvarez:2012wf}. We illustrate the procedure using mainly a signal mass point $(m_{\slep}, m_{\none})=(120,110)\gev$, while presenting results for other masses in Section \ref{sec:conclusion}.
\begin{enumerate}
  \item A leading jet with $p_T>100\gev$ and $|\eta|<2.5$, and $\met>100\gev$.\\
These cuts comply with the ATLAS/CMS \cite{ATLAS:2012zim, CMS:rwa} monojet trigger at 8 TeV. A higher threshold will be used at 14 TeV with more pileup events, in which case, one may need to combine mono-jet events with event samples collected from single-lepton and dilepton triggers, or/and pre-scaled samples. Given the importance of dilepton plus monojet events in both electroweakino and slepton searches, we believe a dedicated trigger should be designed and included in the trigger menu. For this reason, we will use a 100 GeV threshold to explore the LHC discovery potential, while leaving the trigger implementation to experimental experts. We have also tried to increase the jet $p_T$ cut and missing energy cut to both 300 GeV while keeping all other cuts intact. For a typical mass point, (120, 110) GeV, this results in a reduced signal rate to 8\% of that using 100 GeV cuts. Nonetheless, $S/\sqrt{B}$ is only reduced by 25\% because a large missing $E_T$ cut is more efficient killing backgrounds than the signal.
  \item Veto events with a second jet satisfying $p_T>30\gev$ and $|\eta|<4.5$.
  \item No b-tagged jet with $p_T>20\gev$.
  \item A pair of opposite-sign-same-flavor leptons, each of which satisfies $p_T>7\gev$ and $|\eta| <2.5$.
  \item The reconstructed $m_{\tau\tau} > 150\gev$. \\This cut is used to eliminate the large $Z(\rightarrow \tau\tau)$+jet background. The two $\tau$'s momenta are reconstructed using the collinear assumption \cite{Ellis:1987xu, higgsino}. Then we obtain the $Z$ peak of the $Z$+jet background, and the signal and the other backgrounds are largely flat. Since the signal is not populating the smaller $m_{\tau\tau}$ region, we simple use a lower cut of $150\gev$. By doing so, we only lose a small fraction of the signal (and other background) events (Fig.~\ref{fig:mtautau}).
  \item Upper cuts on the lepton $p_T$'s. \\Leptons in signal events are concentrated in the region just above the acceptance cut, as shown in Fig.~\ref{fig:pt} for signal masses (120, 110) GeV, while leptons in $t\bar t$, $j\ell\ell\nu\nu$ and single-top spread across a much larger region. Cutting on the leading lepton $p_T^{l1}<40\gev$ and subleading lepton $p_T^{l2}<30\gev$, we remove $\sim 80\%$ of  the $t\bar t$, $jll\nu\nu$ and single top backgrounds, while keeping $\sim 75\%$ of the signal events. Although the cut is not efficient for the $Z$+jet background, which has a very similar distribution to the signal, it boosts $S/B$ from 0.036 to 0.14, and increases $S/\sqrt{B}$ by a factor of $\sim 1.7$ -- see Table \ref{tab:xsecs} in the next section. This means a 120GeV/110 GeV slepton/LSP can be discovered at a 5.1$\sigma$ level with 100 fb${}^{-1}$ data. Note the signal $p_T$ distribution depends on the mass splitting, therefore, we will need to adjust this cut to optimize the significance, which means a scan of the cut is needed when the mass splitting is unknown. In the following analysis, we will consider 3 mass splittings, 5 GeV, 10 GeV, and 20 GeV. The corresponding $p_T$ cuts for the leading/subleading leptons are chosen as 25 GeV/15 GeV, 40 GeV/30 GeV and 80 GeV/60 GeV respectively. For 20 GeV splittings, these cuts only cause a minor increase in $S/B$ due to the large overlap between the signal and the backgrounds.
  \item An upper cut on dilepton $\mttwo$ -- see the discussion below.
\end{enumerate}

Although the cuts on lepton $p_T$'s are useful for reducing the backgrounds, it is not a direct measure of the small mass difference between $\slep$ and $\none$.  In particular, if the lepton $p_T$ acceptance cut is higher, or if we consider sleptons produced from heavy particle decays, the lepton $p_T$ distribution may shift to higher values and the cut will become less efficient. It also ceases to increase $S/\sqrt{B}$ when the mass splitting is $\gtrsim 20\gev$. A more direct measure of the mass difference will be favorable. In the case of qusi-degenerate Higgsinos \cite{higgsino, Baer:2014kya}, such a variable is provided by the dilepton invariant mass because the two leptons tend to come from the same decay. For slepton pair production, this is no longer the case because the two leptons necessarily come from two different decay chains. Their invariant mass is then approximately determined by their momentum. This is shown in Fig.~\ref{fig:variables} (b) in the Appendix, where we see the dilepton invariant mass distribution of the signal significantly overlaps with the backgrounds. For completeness, in the Appendix we also examine the variable, $\Delta\phi(\ell^1, p_T^{\text{miss}})$, {\it i.e.}, the difference in azimuthal angle between the leading lepton and the missing transverse momentum, which turns out to be not very useful either.  

A good measure of the mass difference is provided by the variable $m_{T2}$ \cite{Lester:1999tx, Barr:2003rg}. The modern definition of the variable is given in Ref.~\cite{ch}, for an event with two decay chains that both end with a mother particle decaying to an invisible daughter particle and a visible particle. The two mother particles' masses are assumed to be equal, so are the two daughter particles' masses, and the two daughter particles are assumed to be the only invisible particles in the event. For a given (trial) mass of the daughter particle, $\mu$, we then define $m_{T2} (\mu)$ as the minimum mother particle mass that is consistent with the measured kinematics including the visible particles' momenta and the transverse missing momentum. We see that this definition applies perfectly to the case of slepton decays since the two sleptons have the same mass, so do the two LSPs. Although $m_{\none}$ is unknown, we can evaluate $m_{T2}-\mu$ using an arbitrary $\mu$ as a trial $m_{\none}$ \footnote{But within the ballpark of masses we are interested in. We have chosen to present the results for a trial LSP mass of 150 GeV, and verified that changing the trial mass to 200 GeV only slightly changes the final results.}, which is,  to a good approximation, still bound by the real mass difference between the two particles.

\begin{figure}
\begin{center}
\includegraphics[width=0.7\textwidth]{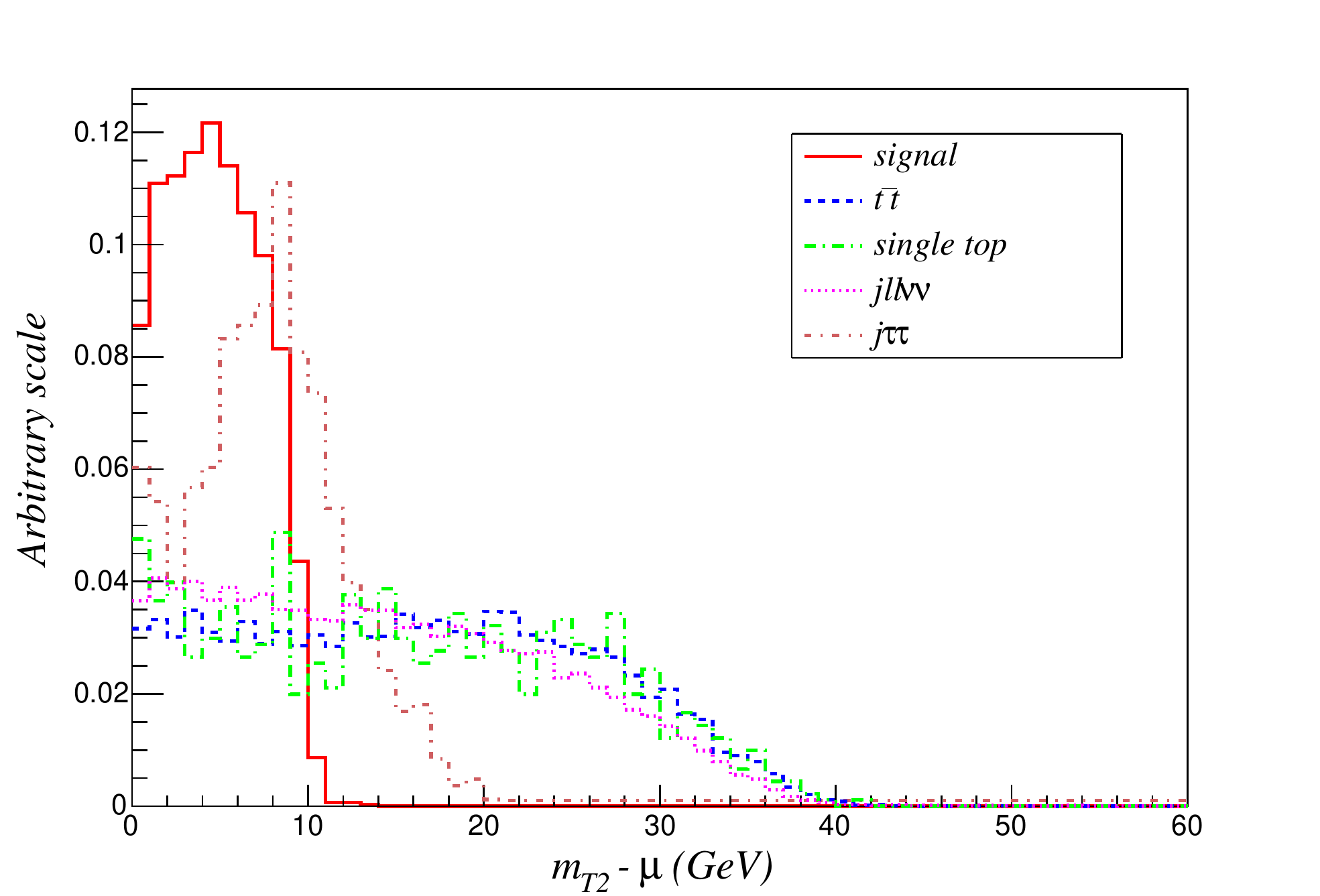} 
\end{center}
\caption{Dilepton $m_{T2}$ distributions, normalized to the same area. The trial mass, $\mu$, for $\none$ is fixed to $120\gev$. Events included in this figure have passed cuts 1-5.}
\label{fig:mt2}
\end{figure}

We show the $m_{T2}$ distributions in Fig.~\ref{fig:mt2}, for signal events from a 120 GeV $\slep$ decaying to a 110 GeV $\none$, and the major SM backgrounds. It is seen that most of the signal events are located below 10 GeV as expected. The distributions for $t\bar t$, $WW$+jets are largely set by the mass difference between the $W$ boson and the neutrino, although it is also shaped by the lepton $p_T$ cuts we have applied. Single-top background has a similar distribution because one of the lepton also comes from a $W$ decay, although we do not have a good understanding why it is so similar to $t\bar t$. The distribution of $Z$+jet is more problematic because it is concentrated on a low mass difference region between 0 and 20 GeV. A cut of $m_{T2}-\mu < 10\gev$ removes $\sim 30 \%$ of $Z$+jet events. For larger signal mass splittings, a larger window in $m_{T2}-\mu$ is needed and more $Z$+jet will be included. Fortunately, $Z$+jet is a minor background once a $m_{\tau\tau}$ cut is imposed. The $m_{T2}-\mu < 10\gev$ cut increases $S/B$ further to 0.37 and $S/\sqrt{B}$ to 8.1 for 120GeV/110GeV $\slep/\none$ with 100 fb${}^{-1}$ data.

\section{LHC reach}
\label{sec:results}
In this section, we vary the slepton mass from 120 GeV to 200 GeV and estimate the LHC reach at 14 TeV for 3 mass splittings, 5, 10 and 20 GeV. For the same mass splitting, we fix the lepton $p_T$ cuts and the cut on $m_{T2}$, as given in Table~\ref{tab:xsecs}, where we show the signal and the background cross sections after each cut.

\begin{table}[t!]
\begin{center}
\begin{tabular}{|c|c|c|c|c||c|c|c|}
\hline
&\multicolumn{7}{|c|}{$\sigma$(fb) at 14 TeV}\\
\hline
& $t\bar t$ (dilep) & single $t$ (lept)&$j\ell\ell\nu\nu$ & $j\tau\tau$ & $\slep$ (120,115) & $\slep$ (120,110) &$\slep$ (120,100)
\\
\hline
before cuts& 19400 & 13080 & 4121 & 1750 & 31.3  & 31.3 & 31.3 
\\ 
\hline
$p_T^j,\,  \slashed{E}_{T}>100$& 6195  & 4121  & 141 & 881 & 18.7 & 18.0 & 16.9
\\ \hline
second jet veto & 598 & 648 & 54.5 & 459 & 9.58 & 8.92 & 7.63
\\
\hline
$b$-jet veto & 153 & 393 & 53.6 & 453 & 9.36 & 8.71 & 7.45
\\
\hline
isolated OSSF leptons & 38.6 & 4.28 & 20.1 & 47.4 & 1.31 & 2.69 & 3.40
\\
\hline
$m_{\tau\tau}>150$&38.1 & 4.25 & 19.6 & 3.53 & 1.19 & 2.36 & 3.14
\\
\hline\hline
$p_T^{l1}<80,p_T^{l2}<60$ & 25.4 & 2.52& 10.4 & 2.93 & -& - &2.61
\\
\hline
$p_T^{l1}<40,p_T^{l2}<30$ & 7.43 & 0.722 & 2.71 & 1.80 & - & 1.81 & -
\\ 
\hline
$p_T^{l1}<25,p_T^{l2}<15$ & 1.02 & 0.114 &  0.457 & 0.846 & 0.844 & - &-
\\
\hline\hline
$m_{T2}-\mu < 20$ & 12.8 & 1.25 & 5.78 & 2.66 & - & - & 2.59
\\
\hline
$m_{T2}-\mu < 10$ &2.31 & 0.239 & 1.02 & 1.3 & - & 1.79 & -
\\
\hline
$m_{T2}-\mu < 5$ & 0.252 & 0.0279 & 0.121 & 0.389  &  0.825 & - & -\\
\hline
\end{tabular}
\end{center}
\caption{Cross sections (in fb) after each cut, for the major backgrounds, and the signal for two generations of left-handed sleptons with degenerate masses, $m_{\slep}=120\gev$, and three mass splittings, 5 GeV, 10 GeV and 20 GeV. Different lepton $p_T$ cuts and $m_{T2}$ cuts are used for different mass splittings. The unit for all masses and momenta is GeV. The cross sections in the row ``before cuts'' are calculated with Madgraph at tree level. \footnote{The cross sections are after generation cuts: a jet $p_T > 80 \gev$ cut is used for the signal ($j\slep\slep$), $j\ell\ell\nu\nu$ and $j\tau\tau$; a missing $E_T > 80\gev$ cut is used for all backgrounds; a lepton $p_T > 5\gev$ cut is also used for $j\ell\ell\nu\nu$. }} 
\label{tab:xsecs}
\end{table}

\begin{figure}
\begin{center}
\begin{tabular}{cc}
  \includegraphics[width=0.4\textwidth]{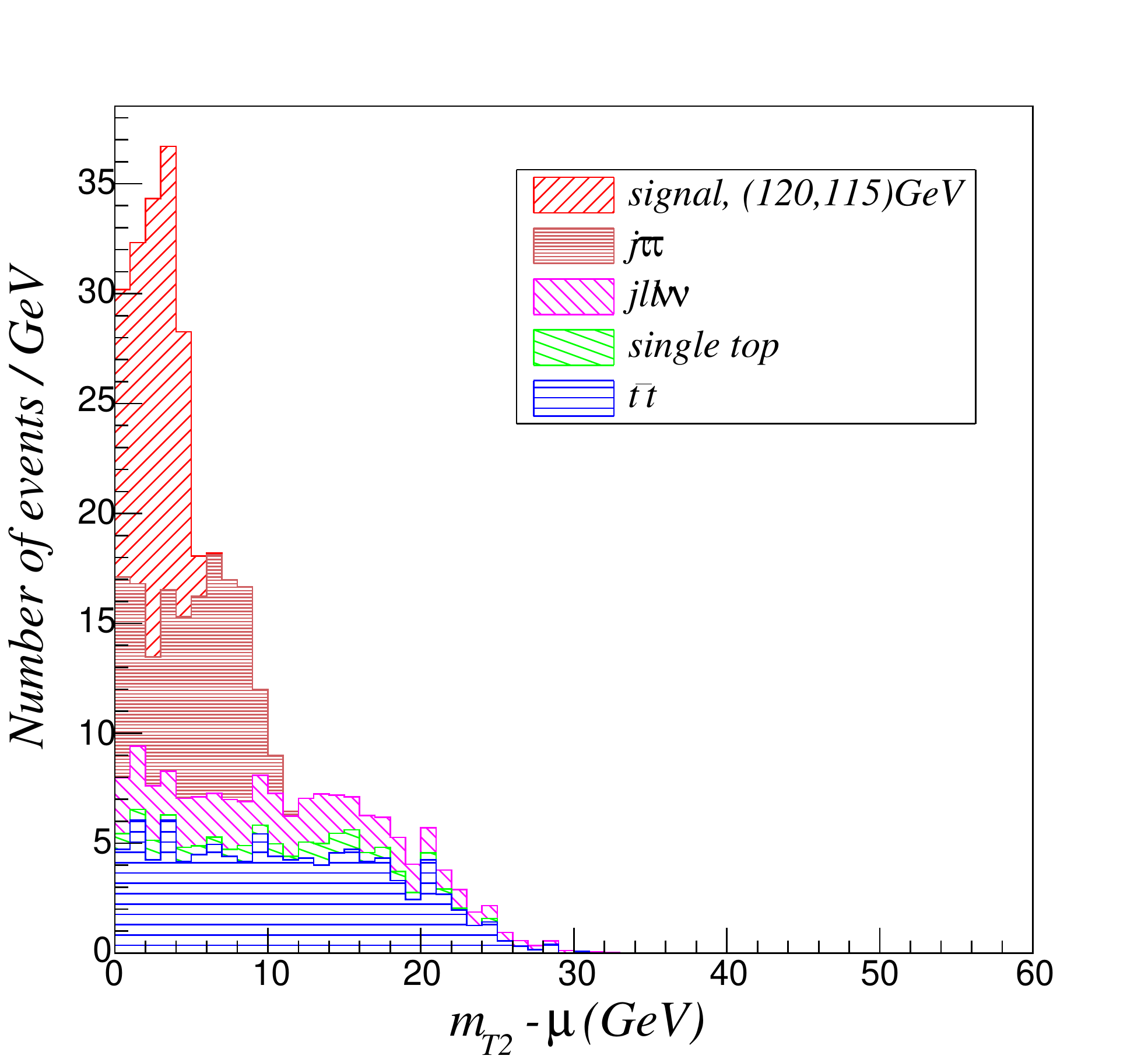} &
  \includegraphics[width=0.4\textwidth]{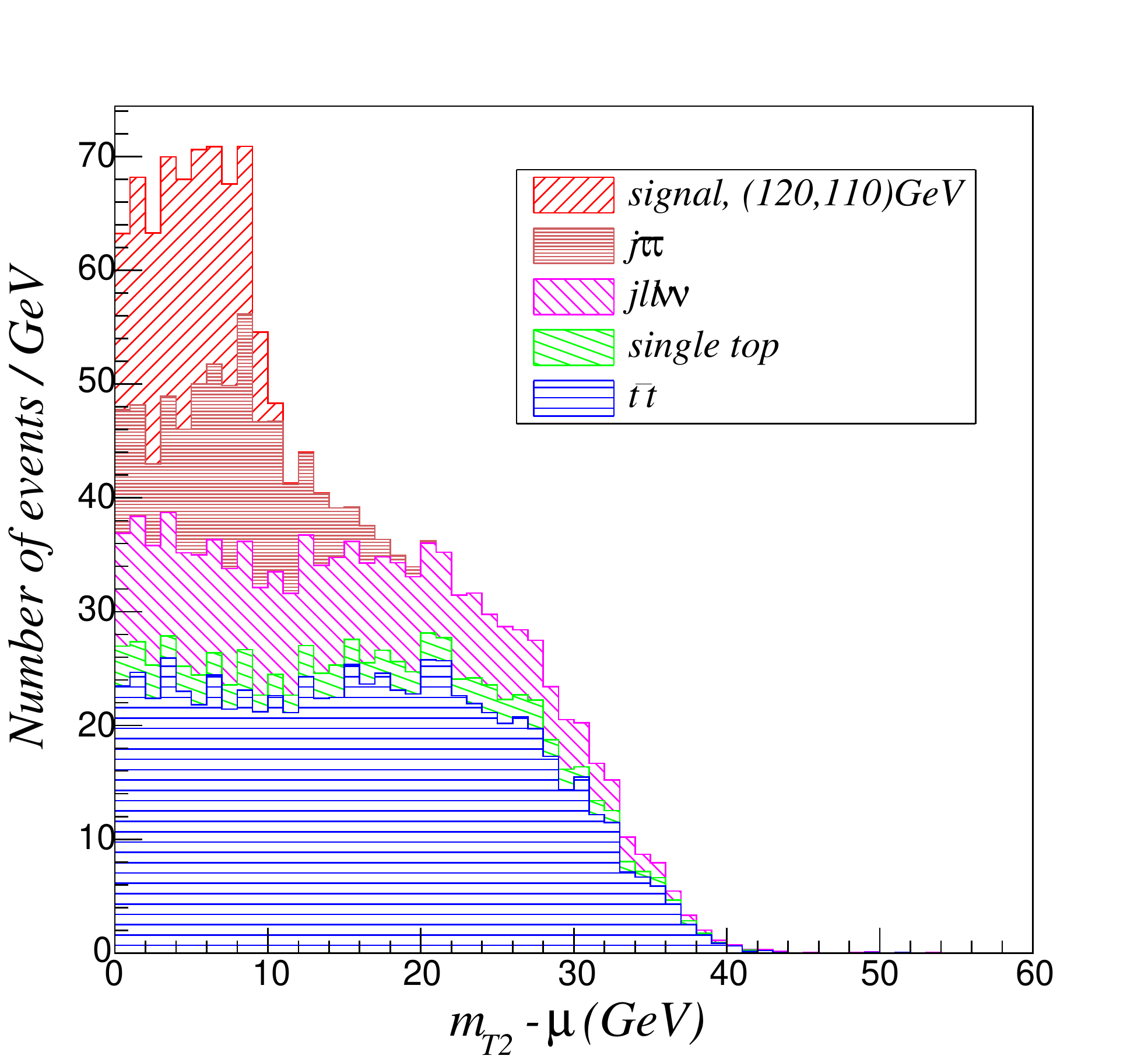} 
\end{tabular}
  \includegraphics[width=0.4\textwidth]{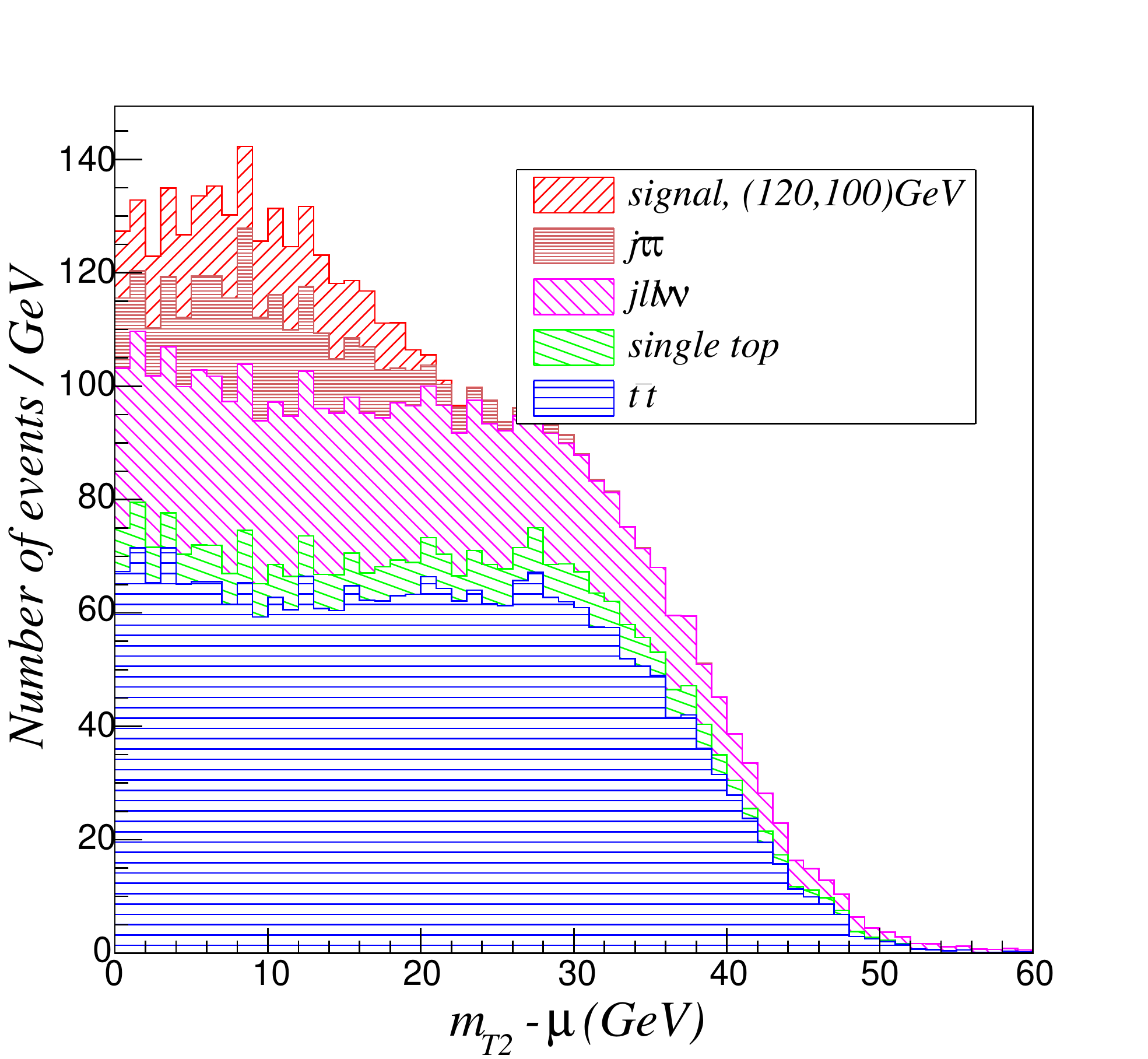}
\end{center}    
\caption{The stacked $m_{T2}$ distributions after all other cuts, for several different mass points. The signal events come from left-handed sleptons of the first two generation (with degenerate masses). The trial mass $\mu$ is fixed to $120\gev$. The number of events correspond to the 14 TeV LHC with 100 fb${}^{-1}$ integrated luminosity.}
\label{fig:mt2-stacked}
\end{figure}

From Table \ref{tab:xsecs}, we see that for the same slepton mass, with a larger mass splitting, we have more signal events with two leptons detected, and after all cuts, more events within the $m_{T2}$ window. However, a smaller mass splitting allows us to use more stringent cuts on the lepton $p_T$ and also a smaller $m_{T2}$ window. Eventually, we obtain a larger $S/B$ and a better significance for a 5 GeV mass splitting than a 20 GeV splitting. This is illustrated in Fig.~\ref{fig:mt2-stacked}, where we show the stacked $m_{T2}$ distributions, assuming an integrated luminosity of 100 fb${}^{-1}$ at the 14 TeV LHC. The signal events include left-handed selectrons and smuons with degenerate masses. For the three mass splittings we consider, For slepton mass of 120 GeV and the three mass splittings, we obtain $S/\sqrt{B}$ of 9.3, 8.1 and 5.4, and $S/B$ of 1.04, 0.37 and 0.12, respectively, by using a cut on $m_{T2}-m_N$ of 5, 10, and 20 GeV. Of course, if the mass splitting is even smaller, we will not be able to collect enough events and the significance will diminish. For example, a 2 GeV mass splitting results in a 0.17 fb effective cross section for detecting two leptons with $p_T > 7\gev$ and a less than 3$\sigma$ significance after imposing an $m_{T2} < 2\gev$ cut. The leptons are also dominantly very close to the threshold and a more careful treatment of the lepton resolution and acceptance may be needed to obtain the precise reach.

\begin{figure}
\begin{center}
\begin{tabular}{cc}
  \includegraphics[width=0.48\textwidth]{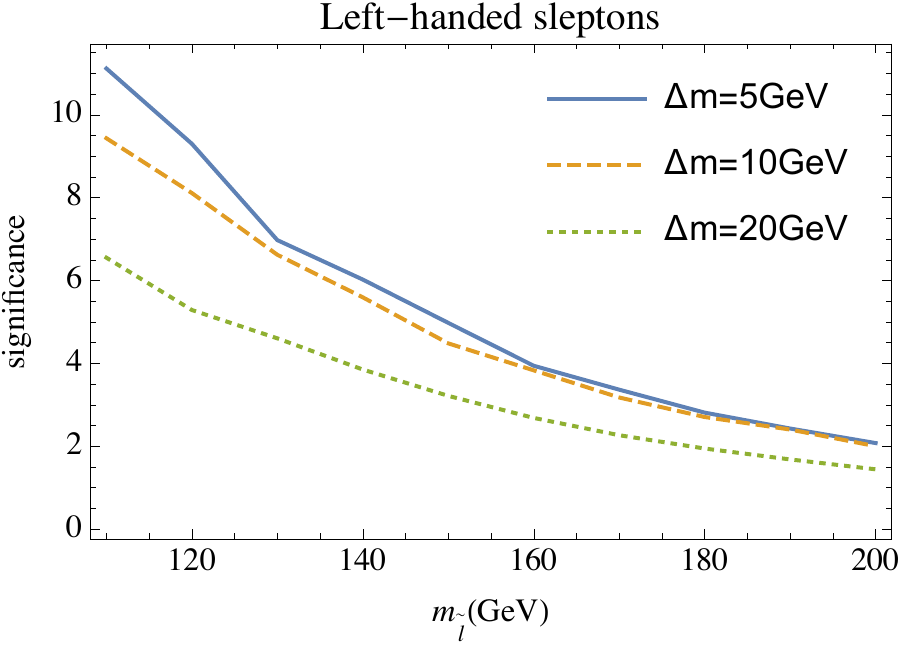} &
  \includegraphics[width=0.48\textwidth]{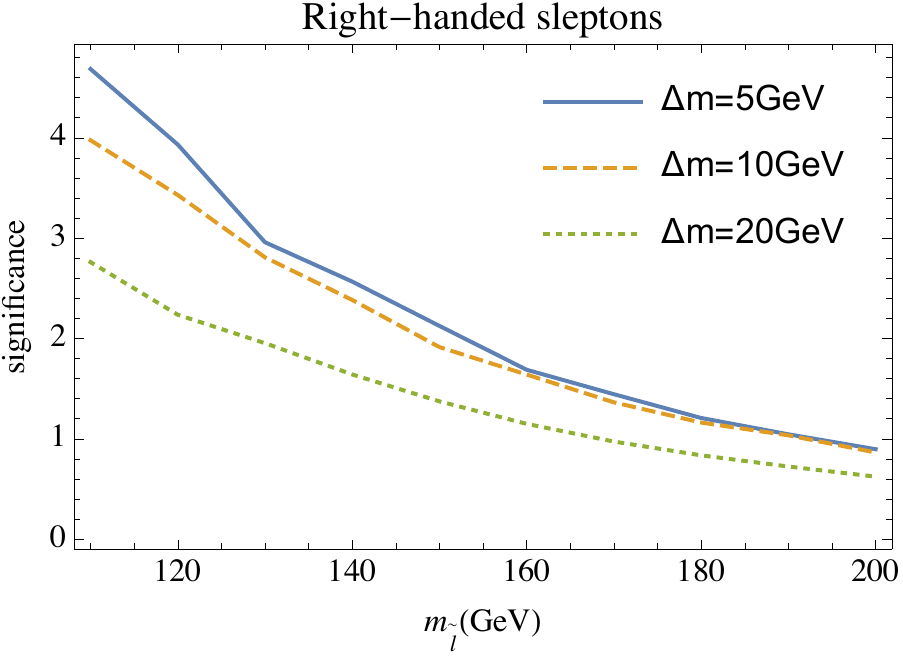} 
\end{tabular}  
\end{center}    
\caption{The statistical significance ($S/\sqrt{B}$) after all cuts, as a function of the slepton mass, for three mass splittings (denoted $\Delta m$). An integrated luminosity of 100 fb${}^{-1}$ at LHC 14 is assumed. Left: left-handed slepton; right: right-handed slepton. Two generations of sleptons (selectons and smuons) of degenerate masses are included.}
\label{fig:reach}
\end{figure}

In Fig.~\ref{fig:reach}, we show $S/\sqrt{B}$ as a function of the slepton mass, for the three mass splittings and for both left-handed and right-handed leptons. There is almost no difference between left-handed and right-handed sleptons in the kinematics. Therefore, the difference in the reach is only caused by the difference in the production cross sections. When producing Fig.~\ref{fig:reach}, we have only included statistical uncertainties and emphasize that systematic errors will be important for higher masses with small signal-background ratios.
     
\section{Discussions}
\label{sec:conclusion}

In Section ~\ref{sec:results}, we quoted our estimates for the LHC discovery limits, which in several ways are conservative. First, we did not fully optimize all kinematic cuts due to limitation in computational powers. Second, we assumed a b-tagging rate of 0.7. Due to the large $t\bar t$ background, a more aggressive b-tagging is beneficial. For example, assuming a b-tagging efficiency of 0.85 and fake rate of 0.1 \cite{Chatrchyan:2012jua}, we can eliminate 50\% more $t\bar t$ events and keep 90\% of the signal. Third, we may have more data than assumed: we used the leading order cross sections, which will be enhanced at NLO; we assumed 100 fb${}^{-1}$ integrated luminosity, while we expect in total more than 300 fb${}^{-1}$ for the LHC and $\sim 3000$ fb${}^{-1}$ for the high luminosity(HL)-LHC, at each of the two experiments. Nevertheless, we see that we are able to reach a $5\sigma$ discovery for $\sim 150$ GeV ($\sim 110$ GeV)  left (right)-handed sleptons. Simply scaling the significance by integrated luminosity, we are able to reach $200\gev$ even for right-handed sleptons at HL-LHC.

Although we focus on slepton searches in this article, the method may be used in searches of other SUSY particles in a compressed spectrum. In order to calculate $m_{T2}$ which is bound by the mass splitting, both visible particles from the two decay chains need to be detected. The visible particles are soft, therefore, they cannot be jets which have large combinatorial backgrounds. These facts limit the use of the method. However, it may be used in cases where visible particles from both decay chains are needed anyway to eliminate large SM backgrounds to signal events with one visible particle lost. Another possible application is in models with gauge mediated supersymmetry breaking with a gravitino LSP, where two photons are produced and play a similar role of the leptons from slepton decay. Moreover, we did not study situations where the sleptons (or other particles as the NLSP) are decay products of much heavier particles. In that case, a large missing $E_T$ is expected and the leptons may be more energetic. Other variables such as the lepton $p_T$ may not be useful, but the $m_{T2}$ distribution will still be bound by the small mass splitting. This feature is unique and cannot be replicated with simple kinematic variables. The reason is, as pointed out in Ref.~\cite{ch}, $\mttwo$ is a root of a 12th order polynomial equation, {\it i.e.}, a complicated function of the 4-momenta of the visible particles and the missing momentum. A large missing $E_T$ cut is commonly used in SUSY searches, so does a large $m_{T2}$ cut. They are usually considered correlated quantities that provide similar information. Here, we emphasize that a large missing $E_T$ cut used simultaneously with a {\it small} $m_{T2}$ as an upper cut might become a crucial criterion that leads to the discovery of supersymmetry.

{\it Note added:} while this work was being completed, we noticed Ref.~\cite{Dutta:2014jda} appeared, which studies compressed sleptons produced in vector boson fusion (VBF) processes. For the same range of masses (115-135 GeV) and mass splittings (5-15 GeV), to obtain a similar significance ($3-6\sigma$), 30 times more data is needed in VBF processes than direct pair production using the method in this article. 
\section*{Acknowledgments}

ZH is supported in part by the US Department of Energy under 
contract NO DE-FG02-96ER40969 and DE-FG02-13ER41986. YL is supported in part by National Science Foundation of China under grant No.~11275009.

\appendix
\section{Comparison to other variables.}
\label{app:comparison}
On the one hand, a small mass splitting lowers the signal acceptance rate, and makes it subject to contamination from large SM backgrounds. On the other hand, an extremely small mass splitting is not usually present in SM processes, which potentially can be used to distinguish signals in compressed spectra from the backgrounds. Besides the variables we have used in our analysis, a number of others have been proposed to capitalize on the small mass splitting. 

\begin{figure}
\begin{center}
\begin{tabular}{cc}
\includegraphics[width=0.5\textwidth]{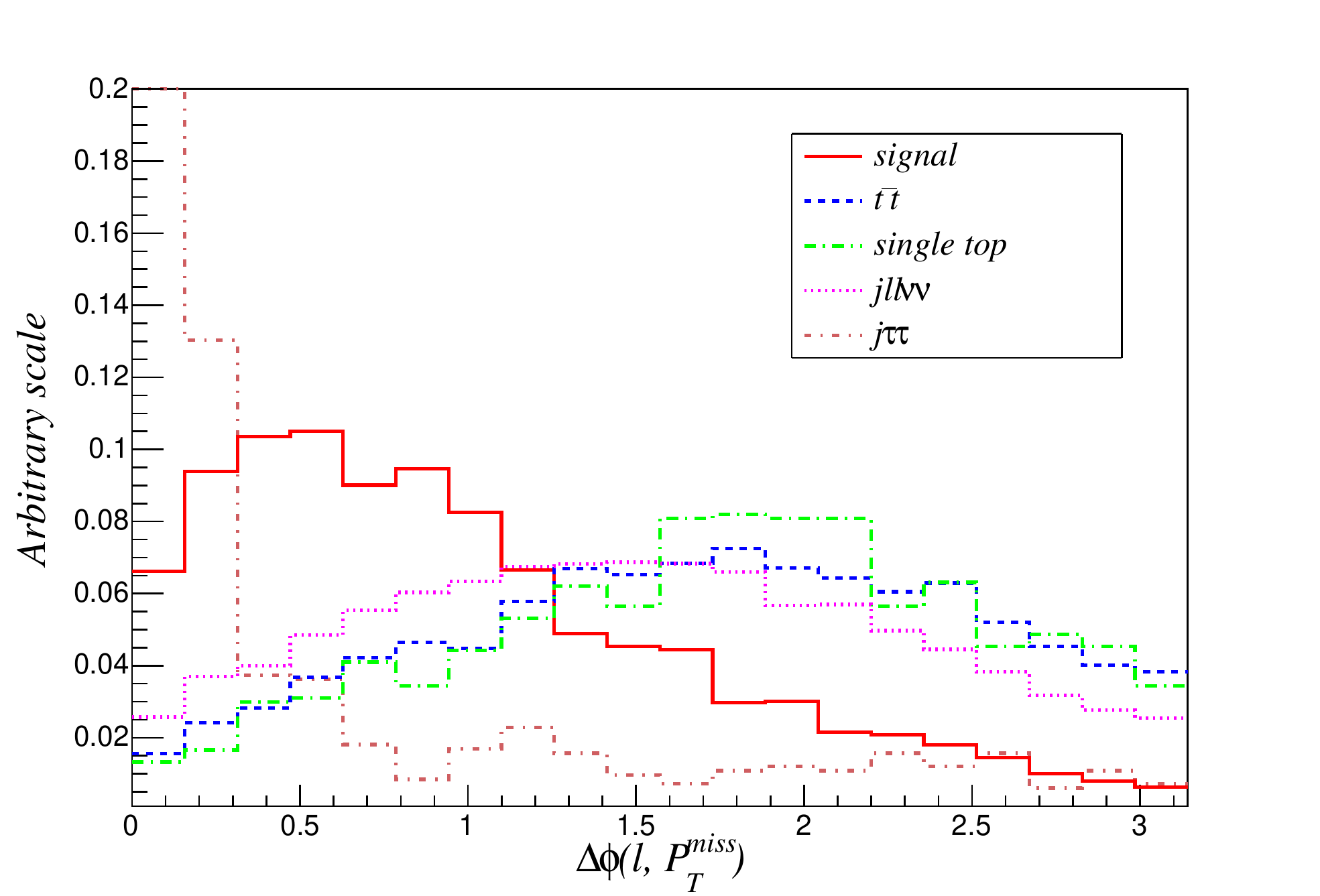}& \includegraphics[width=0.5\textwidth]{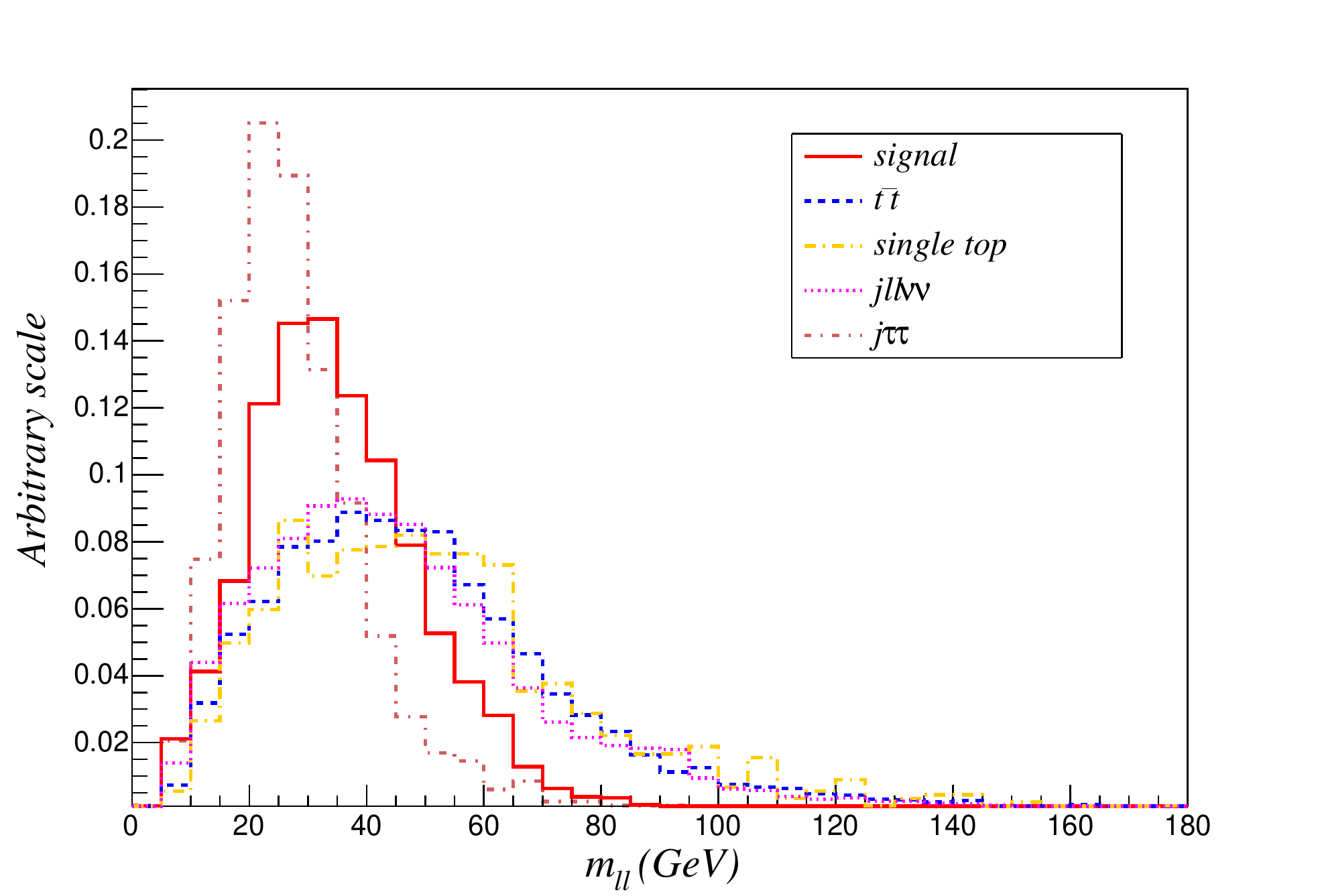} \\
(a)&(b)
\end{tabular}
\end{center}
\caption{Other variables sensitive to a small mass splitting. (a): the $\phi$ angle difference between the leading lepton and the missing momentum (the first bin of $Z$+jet, which extends to 0.61, is truncated for better illustration); (b) The dilepton invariant mass. The distributions are after all cuts described in the main text except for the $m_{T2}$ cut.}
\label{fig:variables}
\end{figure}

Because of the presence of the  monojet, the two sleptons are boosted in the direction opposite to the jet, which makes the decay products of the two sleptons to some extent close to one other. Therefore, we expect the angles between the leptons and the missing transverse momentum to be small for the signal (as well as for the $Z$+jet background). This is manifest in, for example, the $\phi$ angle difference between the leading lepton (denoted $\ell^1$) and the missing momentum, as shown in Fig.~\ref{fig:variables} (a). We have used the (120, 110) GeV signal mass point for Fig.~\ref{fig:variables}, and used the same cuts as in the main text, except for the final $m_{T2}$ cut. This variable is useful, but does not perform as well as the $m_{T2}$ cut. For example, consider the signal and the largest background, $t\bar t$: the best improvement in $S/\sqrt{B}$ occurs when we cut at $\delta\phi(\ell^1, p_T^{\text{miss}}) < 1.1$, which gives us a signal ($t\bar t$ background) efficiency of 0.64 (0.24). For comparison, the $m_{T2} < 10\gev$ cut retains 99\% of the signal events and 31\% of $t\bar t$ events, boost $S/\sqrt{B}$ by a factor of 1.8. Nevertheless, this variable may be important in the case when one of the visible particles is undetected and the $m_{T2}$ variable is not calculable, such as in a sbottom search \cite{Alvarez:2012wf}.

In a search of quasi-degenerate Higgsinos \cite{higgsino, Baer:2014kya}, a cut on the dilepton invariant mass is used to separate the signal from the backgrounds. In that case, the two leptons in  an event can come either from the same or two different decay chains. However, since the mass differences are small, it is difficult to boost both decay chains such that two leptons from different decay chains both pass the acceptance $p_T$ cut. On the other hand, for dileptons from the same particle decay, {\it i.e.}, $\ntwo$ decaying to $\none$ through an off shell $Z$, only one boost is needed since the two leptons have to be close to each other because their invariant mass is small (bound by the $\ntwo$-$\none$ mass difference). As a result, the majority of lepton pairs come from $\ntwo$ to $\none$ decays through an offshell $Z$ boson and a dilepton invariant mass is useful. In our case, the two leptons come from two different decay chains and their invariant mass has a more similar distribution to the backgrounds, as seen in Fig.~\ref{fig:variables} (b). We may also use this variable to increase $S/B$, but it is not as good as a $m_{T2}$ cut.

Although these variables are more or less correlated, they each contain their own information. Therefore, we might be able to obtain a more efficient use of these variables by combining them in a multivariate analysis. This is an interesting approach but beyond the scope of this article.

\bibliography{slep}
\bibliographystyle{JHEP}

\end{document}